\documentclass[twocolumn,showpacs,preprintnumbers,amsmath,amssymb]{revtex4}

\usepackage{graphicx}
\usepackage{dcolumn}
\usepackage{bm}

\begin{document}

\preprint{APS/123-QED}

\title{
Two-orbital model explains the higher transition temperature of the
single-layer Hg-cuprate superconducter compared to that of the 
La-cuprate superconductor
}

\author{Hirofumi Sakakibara$^1$}
\author{Hidetomo Usui$^1$}
\author{Kazuhiko Kuroki$^{1,4}$\onlinecite{corres}}
\author{Ryotaro Arita$^{2,4,5}$}
\author{Hideo Aoki$^{3,4}$}

\affiliation{$\rm ^1$Department of Applied Physics and Chemistry, The University of Electro-Communications, Chofu, Tokyo 182-8585, Japan}
\affiliation{$\rm ^2$Department of Applied Physics, The University of Tokyo, Hongo, Tokyo 113-8656, Japan}
\affiliation{$\rm ^3$Department of Physics, The University of Tokyo, Hongo, Tokyo 113-0033, Japan}
\affiliation{$^4$ JST, TRIP, Sanbancho, Chiyoda, Tokyo 102-0075, Japan}
\affiliation{$^5$ JST, CREST, Hongo, Tokyo 113-8656, Japan}

\date{\today}

\begin{abstract}
In order to explore the reason why the single-layered cuprates, 
La$_{2-x}$(Sr/Ba)$_x$CuO$_4$ ($T_c\simeq$ 40K) 
and HgBa$_2$CuO$_{4+\delta}$ ($T_c\simeq$ 90K), have 
such a significant difference in $T_c$, we study 
a two-orbital model that incorporates the $d_{z^2}$ orbital 
on top of the $d_{x^2-y^2}$ orbital.  
It is found, with the fluctuation exchange approximation, 
that the 
$d_{z^2}$ orbital contribution to the Fermi surface, which is 
stronger in the La system, works against $d$-wave superconductivity, 
thereby dominating over the effect of the Fermi surface shape. 
The result resolves the long-standing contradiction between 
the theoretical results on Hubbard-type models 
and the experimental material dependence of $T_c$ in the cuprates. 
\end{abstract}

\pacs{74.20.-z, 74.62.Bf, 74.72.-h}
\maketitle
The physics of high-$T_c$ superconductivity, despite its long 
history, harbors rich problems which are still open.   
Specifically, given the seminal discovery of 
the iron-based superconductors\cite{Hosono}  and their 
striking material dependence of $T_c$\cite{Lee}, it should be 
important as well as intriguing to have a fresh look at  
the cuprates, which still have the highest $T_c$ to date, to 
understand 
their material dependence of the $T_c$.  
One of the basic problems is the 
significant difference in $T_c$ within the single-layered materials,
i.e., La$_{2-x}$(Sr/Ba)$_x$CuO$_4$ with a maximum $T_c$ of about 40K 
versus HgBa$_2$CuO$_{4+\delta}$ with a $T_c \simeq 90$K. 
Phenomenologically, it has been recognized that the materials 
with $T_c\sim 100$K tend to have ``round'' Fermi surfaces,
while the Fermi surface of the 
La system is closer to a square shape which implies a relatively 
better nesting\cite{Pavarini,Tanaka}.

Conventionally, the materials with a rounded Fermi surface have been 
modeled by a single-band model with large second ($t_2(>0)$) and 
third ($t_3(<0)$) neighbor hopping integrals,  
while the ``low-$T_c$'' La system has been considered to have 
smaller $t_2, t_3$. 
This, however, has brought about a contradiction between 
theories and experiments. 
Namely, while some phenomenological\cite{Moriya} 
and $t$-$J$ model\cite{Shih,Prelovsek} studies 
give a tendency consistent with the experiments, 
a number of many-body 
approaches for the Hubbard-type models with 
realistic values of on-site $U$ show suppression of superconductivity for 
large $t_2>0$ and/or $t_3<0$, 
as we shall indeed confirm below\cite{Scalapino}.

To resolve this discrepancy,  here we consider 
a two-orbital model that explicitly incorporates the $d_{z^2}$ orbital 
on top of the $d_{x^2-y^2}$ orbital.  The former component 
has in fact 
a significant contribution to the Fermi surface in the 
La system. 
We shall show that 
the key parameter that determines $T_c$ is the 
energy level difference between the $d_{x^2-y^2}$ and $d_{z^2}$ orbitals, 
i.e.,  the weaker the $d_{z^2}$ contribution to the 
Fermi surface, the better for $d$-wave superconductivity, 
where 
a weaker contribution of the $d_{z^2}$ results in a rounded 
Fermi surface (which in itself is not desirable for superconductivity), 
but it is the ``single-orbital nature'' that favors a higher $T_c$ dominating 
over the effect of the Fermi surface shape for the La system.  
\begin{table}[!b]
\caption{Hopping integrals within the $d_{x^2-y^2}$ orbital  
for the single and two orbital models, and $\Delta E\equiv E_{x^2-y^2}-E_{z^2}$.}
\label{hop-para}
\begin{tabular}{ c| l l l l}
\hline
 & 1-orbital & & \hspace{1.0cm} 2-orbital & \\
 &  La & Hg  &\hspace{1.0cm}  La & Hg\\ \hline
$t_1 \rm{[eV]}$ & -0.444 & -0.453 & \hspace{1.0cm} -0.471 & -0.456 \\ 
$t_2 \rm{[eV]}$ & 0.0284 & 0.0874 & \hspace{1.0cm} 0.0932 & 0.0993 \\
$t_3 \rm{[eV]}$ & -0.0357 & -0.0825  & \hspace{1.0cm}  -0.0734 & -0.0897 \\
$(|t_2|+|t_3|)/|t_1|$ & 0.14 &  0.37 & \hspace{1.0cm} 0.35 & 0.41 \\  
$\Delta E\rm{[eV]}$ & -  &  - &\hspace{1.0cm}  0.91 & 2.19 \\
\hline
\end{tabular}
\end{table}

\begin{figure}[!t]
\includegraphics[width=7cm]{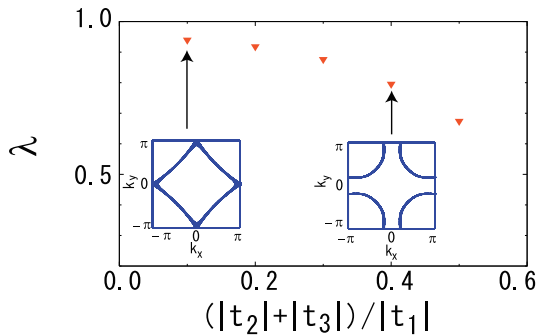}
\caption{FLEX result for the eigenvalue, $\lambda$, of the 
Eliashberg equation for the single-band Hubbard model plotted as 
a function of $(|t_2|+|t_3|)/|t_1|$, where we take $t_2=-t_3 > 0$ 
for $U=6|t_1|$, $T=0.03|t_1|$, and 
the band filling $n=0.85$.  
Fermi surfaces are displayed for two cases (indicated by arrows).}
\label{fig1}
\end{figure}

Let us start with a conventional calculation for the 
single-band Hubbard Hamiltonian, 
$H=\sum_{ij\sigma}t_{ij}c_{i\sigma}^\dagger c_{j\sigma} 
+ U\sum_i n_{i\uparrow}n_{i\downarrow}$.
Here we take the 
nearest-neighbor hopping $-t_1$ ($\simeq 0.4$eV, see table I) 
to be the unit of energy, 
$U=6$, the temperature $T=0.03$, 
and the band filling $n=0.85$ are fixed, 
while we vary $t_2=-t_3$ with $t_2>0$.  
We then apply the fluctuation exchange 
approximation (FLEX)\cite{Bickers,Dahm} 
to solve the linearized Eliashberg equation.   
$T_c$ is the temperature at which the eigenvalue $\lambda$ 
of the Eliashberg equation reaches unity, 
so $\lambda$ at a fixed temperature can be used as a measure
for the strength of the superconducting instability.
We show in Fig.\ref{fig1} $\lambda$ 
as a function of $(|t_2|+|t_3|)/|t_1|$($=2|t_2|/|t_1|$ here), 
which just confirm that, within the single-band 
model, $\lambda$ (hence $T_c$) monotonically 
decreases with increasing $|t_2|$ and $|t_3|$. 
A calculation with the 
dynamical cluster approximation (DCA) shows that a negative $t_2$ works 
destructively against $d$-wave superconductivity\cite{Maier}, and 
a more realistic DCA calculation that considers the oxygen $p_\sigma$ 
orbitals for the La and Hg cuprates also indicates 
a similar tendency\cite{Kent}.   
As mentioned above, this seems to contradict 
with the experimental results that the materials with 
larger $t_2$ and $t_3$ 
have actually higher $T_c$'s\cite{Pavarini}.

\begin{figure}[!t]
\includegraphics[width=8cm]{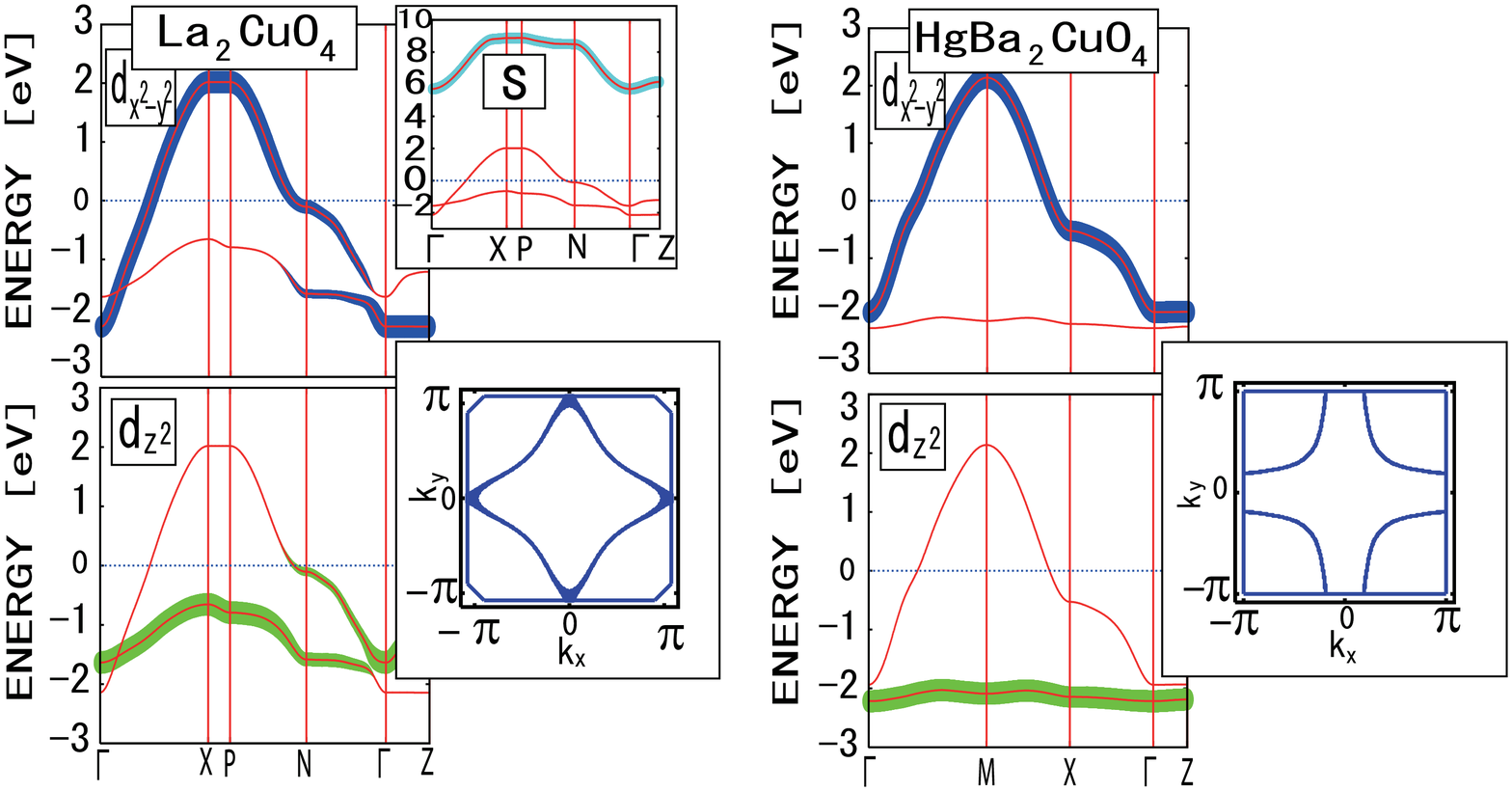}
\caption{The band structure in the two 
($d_{x^2-y^2}$-$d_{z^2}$) orbital model for La$_2$CuO$_4$ (left) and 
HgBa$_2$CuO$_4$ (right). The top (middle) panels 
depict the strength of the $d_{x^2-y^2}$ 
($d_{z^2}$) characters with thickened lines, while the bottom panels the 
 Fermi surfaces (for a total band filling $n=2.85$).  
The inset shows the band 
structure of the three-orbital model (see text) for La system, where the 
the 4s character is indicated.}
\label{fig2}
\end{figure}

To resolve this, we now  introduce the 
$d_{x^2-y^2}$-$d_{z^2}$ two-orbital 
model. 
For the La system, it has long been 
known that a band with a strong $d_{z^2}$ 
character lies rather close to the Fermi energy\cite{Shiraishi,Eto,Freeman}.
More recently, it has been discussed in Refs.\cite{Andersen,Pavarini} 
that the shape of the Fermi surface is determined by the energy level of the 
``axial state'' consisting of a mixture of Cu $d_{z^2}$-O $p_z$ and 
Cu $4s$ orbitals, and 
that the strength of the $d_{z^2}$ contribution 
causes the difference in the 
Fermi surface shape between the La and Hg systems. 
Namely, the $d_{z^2}$ contribution is large in the 
La system making the Fermi surface closer to a square, 
while the contribution is small in the Hg system making the Fermi surface 
more rounded.   In Fig.\ref{fig2}, we show the present, 
first-principles\cite{pwscf} result for 
band structures in the two-orbital model for the La and Hg systems, 
obtained by constructing maximally localized Wannier 
orbitals\cite{MaxLoc}.  
The lattice parameters adopted here are experimentally determined 
ones for the doped materials\cite{La-st,Hg-st}. 
We can here confirm that in the La system 
the main band (usually considered 
to be the ``$d_{x^2-y^2}$ band'') has in fact a strong $d_{z^2}$ character 
on the Fermi surface near the N point, which corresponds to the wave vectors 
 $(\pi,0), (0,\pi)$ in the Brillouin zone of the square lattice.  
The $d_{z^2}$ contribution is seen to ``push up'' 
 the van Hove singularity (vHS) of the main band, resulting in a 
seemingly well nested (square shaped) 
Fermi surface.  In the Hg system, on the other hand, the $d_{z^2}$ band 
stays well away from $E_F$, and consequently the vHS 
is lowered, resulting in  a rounded Fermi surface.

If we estimate in the two-orbital model   
the ratio $(|t_2|+|t_3|)/|t_1|$ 
within the $d_{x^2-y^2}$ orbitals, we get 
0.35 for the La system against 
0.41 for Hg (table I), which are rather close to each other. 
This sharply contrasts with the situation in which the model 
is constrained into a single band. 
There, the Wannier orbital has mainly $d_{x^2-y^2}$ character,
but has ``tails'' with a $d_{z^2}$ character especially for the La system. 
Then the ratio $(|t_2|+|t_3|)/|t_1|$ in the single-orbital model 
reduces to 0.14 for La against 0.37 for Hg (table I), 
which is just the conventional view 
mentioned in the introductory part.  
From this, we can confirm that  it is the $d_{z^2}$ 
contribution that makes the Fermi 
surface in the La system square shaped, while the 
``intrinsic'' Fermi surface of the 
high $T_c$ cuprate family is, as in the Hg system, 
rounded. 

\begin{figure}[!b]
\includegraphics[width=6.5cm]{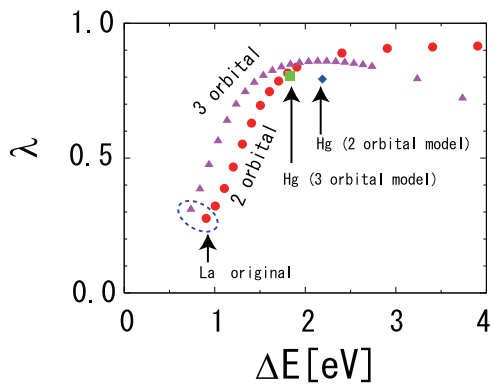}
\caption{The eigenvalue, $\lambda$, of the Eliashberg equation for $d$-wave 
superconductivity is plotted against 
$\Delta E=E_{x^2-y^2}-E_{z^2}$ 
for the two-orbital (red circles) or three-orbital 
(purple triangles) models for La$_2$CuO$_4$. 
Corresponding eigenvalues 
for HgBa$_2$CuO$_4$ are also indicated.
}
\label{fig3}
\end{figure}

Now we come to the superconductivity in the two-orbital model. 
For the electron-electron interactions,
it is widely accepted that the intraorbital $U$ is $7-10t$ 
(with $t\sim 0.45$eV) for the cuprates, so we take $U=3.0$eV. 
The Hund's coupling $J$ (= 
pair-hopping interaction $J'$) is typically $\sim 0.1U$, 
so here we take $J=J'=0.3$eV, which gives the interorbital $U'=U-2J=2.4$eV.
The temperature is fixed at $k_BT=0.01$eV.
As for the band filling (number of electrons/site), 
we concentrate on the total 
$n=2.85$, for which the main band has 0.85.
Here we apply the multiorbital FLEX, as described e.g. in ref.\cite{Kontani}, 
for the three-dimensional lattice 
taking $32\times 32\times 4$ $k$-point meshes and 1024 Matsubara frequencies.
We first focus on the 
La system, and investigate how the 
$d_{z^2}$ orbital affects superconductivity. 
Namely, while 
the on-site energy difference, 
$\Delta E\equiv E_{x^2-y^2}-E_{z^2}$, 
between the two orbitals is 
$\Delta E\simeq 0.9$eV for La$_2$CuO$_4$ 
(table I), we vary the value to probe 
how the Eliashberg 
eigenvalue $\lambda$ for $d$-wave superconductivity behaves. 
The result in Fig.\ref{fig3} shows that 
$\lambda$ is small for the original value of $\Delta E$, but 
rapidly increases with $\Delta E$, until it 
saturates for sufficiently large $\Delta E$. 
Hence the superconductivity turns out to be enhanced 
as the  $d_{z^2}$ band moves away from the main band.  
Note that this occurs despite the 
Fermi surface becoming more rounded with larger $\Delta E$, 
namely, the effect of the 
orbital character (smaller $d_{z^2}$ contribution) 
dominates over the Fermi surface shape effect.  
Conversely, 
the strong $d_{z^2}$ orbital character 
in the Fermi surface around the   
$(\pi,0),(0,\pi)$ works destructively against $d$-wave 
superconductivity. Physically, the reason for this may be explained 
as follows.  
First, although the La system has a better nested Fermi surface, 
we find that the strength of the antiferromagnetic 
spin fluctuations (the spin susceptibility obtained in FLEX) in La is 
only as large as that for Hg.
This is intuitively understandable, since 
the two electrons on nearest-neighbor sites are less constrained 
to have antiparallel spins in order to gain kinetic energy 
when two orbitals are active as in La.
Secondly, $d$-wave pairing 
has a  rough tendency for higher $T_c$ in 
bands that are nearly half filled, whereas the 
$d_{z^2}$ orbital here is nearly full filled.

\begin{figure}[!b]
\includegraphics[width=9.0cm]{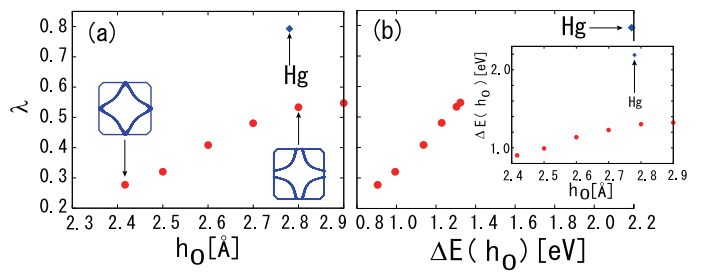}
\caption{The eigenvalue of the Eliashberg equation 
$\lambda$ (red circles) when $h_{\rm O}$(a) or $\Delta E$(b) is varied 
hypothetically in the lattice structure of La$_2$CuO$_4$. 
Blue diamond indicates the eigenvalue of HgBa$_2$CuO$_4$.
Inset in (b) shows the relation  between $h_{\rm O}$ and $\Delta E$. 
}
\label{fig4}
\end{figure}

We now focus on how the lattice structure affects $\Delta E$ and 
hence superconductivity. This is motivated by the fact that 
$\Delta E$ should be controlled 
by the ligand field, hence by the height, $h_{\rm O}$, of the 
apical oxygen above the CuO$_2$ plane\cite{Eto}.  
To single out this effect, 
let us examine the two-orbital model for which 
we increase $h_{\rm O}$ from its 
original value $2.41$\AA\hspace{0.1cm}with other lattice parameters fixed.  
In Fig.\ref{fig4}(a), which plots the eigenvalue of the Eliashberg equation 
as a function of $h_{\rm O}$, we can see that 
$\lambda$ monotonically increases with the height.
As seen from the inset of Fig.\ref{fig4}(b), 
$\Delta E$ is positively correlated with 
$h_{\rm O}$ as expected, and Fig.\ref{fig4}(b) 
confirms that the increase in $\lambda$ 
is due to the increase in $\Delta E$\cite{comment2} .
In these figures, we have also plotted the 
values corresponding to the Hg system  
obtained with the actual lattice structure. 
We can see that, while $h_{\rm O}\simeq 2.8{\rm \AA}$ for Hg 
is larger than $h_{\rm O}\simeq 2.4{\rm \AA}$ for La, 
$\Delta E\simeq 2.2$eV for Hg  is even 
larger than $\Delta E\simeq 1.3$eV, which is the value the La system 
would take for $h_{\rm O}=2.8{\rm \AA}$.
Consequently, $\lambda$ 
for Hg is somewhat larger than 
that for the La system with the same value of $h_{\rm O}$.  
This implies that there are some effects other than the apical oxygen 
height that also enhance $\Delta E$ 
in the Hg system, thereby further favoring 
$d$-wave superconductivity.  In this context, 
the present result reminds us of the so-called 
``Maekawa's plot'', where a positive correlation between $T_c$ 
and the level of the apical oxygen $p_z$ hole was 
observed\cite{Maekawa}.  
Since a higher $p_z$ hole level (i.e., a lower 
$p_z$ electron level) is likely to 
lower $E_{z^2}$, the positive correlation between 
$\Delta E$ and $T_c$ found here is indeed 
consistent with Maekawa's plot.  
It can be considered that in La cuprates, 
a considerable portion of the doped holes go into the apical oxygen 
$p_z$, and this effect is effectively taken into account in our model.
A more detailed study on these issues is 
now under way, and will be discussed in a separate publication.

Finally, let us discuss the effect of Cu $4s$ orbital, 
which is the main component of 
the ``axial state'' discussed in Refs.\cite{Andersen,Pavarini}. 
In the present two-orbital model the $4s$ orbital is 
effectively incorporated in both of the $d_{x^2-y^2}$ and $d_{z^2}$ 
orbitals, i.e., the Wannier orbitals have tails that have the $4s$ character.
In order to make the examination more direct, 
we now consider a three-orbital model that 
explicitly considers the $4s$ orbital.
The band dispersion 
for the La system shown in the 
inset of Fig.\ref{fig2} shows that the $4s$ band lies 
well ($\simeq 7$ eV) above the Fermi level. 
Nonetheless, the $4s$ orbital gives 
an important contribution to the Fermi surface 
in that the ratio 
$(|t_2|+|t_3|)/|t_1|$ within the $d_{x^2-y^2}$ sector 
in the three-orbital model 
takes a much smaller value of 
$0.10$, which should imply that it is the path 
$d_{x^2-y^2}\rightarrow 4s\rightarrow d_{x^2-y^2}$ that gives the 
effectively large $t_2$, $t_3$, 
and hence the 
round Fermi surface, as pointed out 
previously\cite{Andersen,Pavarini}.
In this context, it is worth mentioning that the path 
 $d_{x^2-y^2}\rightarrow d_{z^2}\rightarrow d_{x^2-y^2}$ 
also contributes to $t_2$, $t_3$,  
but has an opposite sign to the $4s$ contribution because the 
$d_{z^2}$ level lies below $d_{x^2-y^2}$, 
while $4s$ above $d_{x^2-y^2}$\cite{Hansmann}.  
So the two contributions to the main band cancel with each other, 
where the cancellation should be strong when the energy 
of the $d_{z^2}$ orbital is high as in La.

We now apply FLEX to the three-orbital model 
varying $\Delta E=E_{x^2-y^2}-E_{z^2}$ 
as in the two-orbital model, where we 
fix the on-site energy difference 
$E_{4s}-E_{z^2}$ 
at its original value.  
We have chosen this 
because a 
similar three-orbital model constructed for Hg (not shown) 
shows that the on-site energy difference between 
the $4s$ and $d_{x^2-y^2}$ orbitals is smaller than in 
the La system by about 1eV, so in the Hg system, both of 
$E_{x^2-y^2}-E_{z^2}$ and 
$E_{4s}-E_{x^2-y^2}$
are smaller by about 1eV, which means that 
the $d_{z^2}$ and $4s$ levels shift roughly in parallel relative 
to $d_{x^2-y^2}$. 
It can be seen in Fig.\ref{fig3} 
that the $\Delta E$ dependence of $\lambda$ in 
the three-orbital model resembles that of the two-orbital model in 
the realistic $\Delta E$ range. (When $\Delta E$ becomes 
unrealistically large, i.e., when $4s$ level is too close to the 
Fermi level, the Fermi surface 
becomes too deformed for superconductivity to be retained.)
We have also calculated the eigenvalue for the Hg system in the 
three-orbital model, and obtained a value very similar to that obtained 
in the two-orbital model, as plotted in Fig.\ref{fig3}.  
If we summarize the three-orbital results, 
while the $4s$ orbital has an important effect on the shape of the 
Fermi surface, this 
can be effectively included in the $d_{x^2-y^2}$ and $d_{z^2}$ 
Wannier orbitals in the two-orbital model 
as far as the FLEX studies are concerned. 
This contrasts with the case of the $d_{z^2}$ orbital, which,  
if effectively included in the $d_{x^2-y^2}$ Wannier orbital to construct 
a single-orbital model, would result in a different result.
This conclusion is natural, since the 
energy difference ($\simeq 1$ eV) between 
$d_{x^2-y^2}$ and $d_{z^2}$ orbitals 
in the La system is smaller than the electron-electron 
interaction, which is why the $d_{z^2}$ orbital has to be explicitly 
considered in a many-body analysis, while the energy difference 
($\simeq 7$ eV) between 
$d_{x^2-y^2}$ and $4s$ orbitals is 
much larger 
than the electron-electron interaction, so that the $4s$ orbital 
can effectively be integrated out 
before the many-body analysis.    So the message  
here is that the 
two-orbital ($d_{x^2-y^2}$-$d_{z^2}$) model suffices 
to discuss the material dependence of the $T_c$ in the cuprates. 
Whether the effect of the 
$d_{z^2}$ orbital can be further incorporated in the on-site $U$ or 
off-site $V$ values (i.e., material-dependent interaction values) 
in an effective, single-band model is a future problem.

To summarize, we have introduced a two-orbital model 
to understand the material dependence of $T_c$  in the 
cuprates.
We have shown that the key parameter 
is the 
energy difference between the $d_{x^2-y^2}$ and $d_{z^2}$ orbitals, 
where the smaller the contribution of the $d_{z^2}$ orbital,  
the better for $d$-wave superconductivity, 
with the orbital-character effect superseding 
the effect of the Fermi surface shape. 
It is intriguing to note that the two high $T_c$ families, 
cuprates and iron pnictides, exhibit material dependence of $T_c$  
that, according to the present study and Ref.\cite{Kuroki}, 
owes to 
the material dependent multiorbital band structures.

In the present view, the Hg cuprate is ``ideal'' 
in that the $d_{z^2}$ band lies far below the Fermi level. 
Nevertheless, there is  
still room for improvement: 
as mentioned in the outset, within single-orbital systems 
higher $T_c$ can be obtained for 
smaller $t_2$ and $t_3$. It 
may be difficult to 
make $t_2$ and $t_3$ smaller in the
cuprates, since they are intrinsically large 
as far as the Cu $4s$ orbital is effective. 
Conversely, we can predict that 
materials with an isolated single band that has smaller $t_2$ and $t_3$ 
should accommodate even higher $T_c$ than the Hg cuprate, 
provided that the electron interaction is similar to those in the cuprates.
 
We wish to acknowledge 
Y. Nohara for the assistance in the band calculation of 
the Hg system. R.A. acknowledges X. Yang and O.K. Andersen for 
fruitful discussions.
The numerical calculations were performed at the Supercomputer Center, 
ISSP, University of Tokyo. This study has been supported by 
Grants-in-Aid for Scientific Research from MEXT of Japan and from JSPS. 
H.U. acknowledges support from JSPS.


\begin{thebibliography}{99}
\bibitem{comment3} The conclusion of the present study is not 
sensitive to the choice of the interaction values. For instance, 
if we raise the interactions to $U=4.5$eV $(\sim 10t )$ and $J=0.45$eV,  
the eigenvalue of the Eliashberg equation 
for the two-orbital model discussed in Fig.\ref{fig3} is only 
slightly modified to $\lambda=0.29$ for La and $0.72$ for Hg.
\bibitem{comment2} Note that, while $\Delta E$ is 
varied ``by hand'' in the Hamiltonian in Fig.\ref{fig3}, 
$\Delta E$  in Fig.\ref{fig4} is made to vary with 
$h_{\rm O}$ of the hypothetical lattice structure.
\end{thebibliography}
\end{document}